\documentclass[
reprint,
superscriptaddress,
amsmath,
amssymb,
aps,prl]{revtex4-1}

\usepackage{graphicx}% Include figure files
\usepackage{dcolumn}% Align table columns on decimal point
\usepackage[
]{changes}
\usepackage{amsthm,amsmath,amssymb}\usepackage{mathrsfs}
\usepackage{bm}% bold math
\usepackage{xcolor}
\usepackage[mathlines]{lineno}% Enable numbering of text and display math
\usepackage[normalem]{ulem}
\usepackage[
			colorlinks = true,
            linkcolor = blue,
            urlcolor  = blue,
            citecolor = blue,
            anchorcolor = blue
            ]{hyperref}         
\newcommand\rxout{\bgroup\markoverwith{\textcolor{red}{\rule[.5ex]{2pt}{.6pt}}}\ULon}

\begin{document}

\title{Observation of higher-order exceptional points in a non-local acoustic metagrating}
\author{Xinsheng Fang}
\affiliation{Institute of Acoustics, School of Physics Science and Engineering, Tongji University, Shanghai 200092, China}
\author{Nikhil JRK Gerard}
\affiliation{Graduate Program in Acoustics, Penn State University, University Park, Pennsylvania 16802, USA}
\author{Zhiling Zhou}
\affiliation{Institute of Acoustics, School of Physics Science and Engineering, Tongji University, Shanghai 200092, China}
\author{Hua Ding}
\author{Nengyin Wang}
\author{Bin Jia}
\affiliation{Institute of Acoustics, School of Physics Science and Engineering, Tongji University, Shanghai 200092, China}
\author{Yuanchen Deng}
\affiliation{Graduate Program in Acoustics, Penn State University, University Park, Pennsylvania 16802, USA}
\author{Xu Wang}
\email{xuwang@tongji.edu.cn}
\affiliation{Institute of Acoustics, School of Physics Science and Engineering, Tongji University, Shanghai 200092, China}
\author{Yun Jing}
\email{yqj5201@psu.edu}
\affiliation{Graduate Program in Acoustics, Penn State University, University Park, Pennsylvania 16802, USA}
\author{Yong Li}
\email{yongli@tongji.edu.cn}
\affiliation{Institute of Acoustics, School of Physics Science and Engineering, Tongji University, Shanghai 200092, China}

\date{\today}
\begin{abstract}
Higher-order exceptional points  have attracted increased attention in recent years due to their enhanced sensitivity and distinct topological features. Here, we show that nonlocal acoustic metagratings that enable precise and simultaneous control over their muliple orders of diffraction, can serve as a robust platform for investigating higher-order exceptional points in free space. The proposed metagratings, not only could advance the fundamental research of arbitrary order exceptional points, but could also empower unconventional free-space wave manipulation for applications related to sensing and extremely asymmetrical wave control.
\end{abstract}

\maketitle
Exceptional points (EPs) are singularities in parameter space, uniquely supported by non-Hermitian systems \cite{Bender1998PRL,Dembowski2001PRL,Heiss2004JPM,moiseyev2011nonhermitian,Peng2014NP,Hodaei2017Nature,Miri2019Science}. EPs for classical waves have been realized by employing gain and/or loss, and unconventional wave behaviors have been demonstrated when the eigenvalues and eigenvectors of the system's Hamiltonian or scattering matrix (S-matrix) simultaneously coalesce. Pivoted on the concept of EPs, new mechanisms for controlling light \cite{Feng2014Science,Feng2014OE,Zhang2019PRL,Huang2015OE,Lin2011PRL} and sound \cite{Zhu2014PRX,Ding2015PRX,Liu2018PRL,Achilleos2017PRB} have been identified. Examples include but are not limited to unidirectional wave propagation \cite{Feng2014OE,Liu2018PRL,Huang2015OE,Lin2011PRL}, coherent perfect absorption ~\cite{Achilleos2017PRB}, and phonon lasing ~\cite{Zhang2018NP}. Early research in non-Hermitian wave physics showed that 2nd order EPs can be realized by using two coupled cavities or waveguides, and such low-order EPs can be leveraged for sensing applications ~\cite{Chen2017Nature,Wang2020NC,Djorwe2019PRApplied}. Subsequent studies, however, emerged to show that a far greater sensitivity can be realized at higher-order EPs ~\cite{Hodaei2017Nature}. These higher-order EPs were made possible by increasing the number of cavities or waveguides in the system ~\cite{Hodaei2017Nature,Ding2015PRX,Wu2021OE,Zhou2018APL}. The major downside of this scheme, however, is that the waves are confined in closed systems and hence limited to 0-D or 1-D wave propagation. EPs for open systems permitting wave propagation in higher-dimensions, on the other hand, would entail richer physics and offer a greater variety of wave functionalities. 

Driven by these prospects, more recent studies have drawn inspiration from the progress in 2D wave functional materials ~\cite{Li2013SR,Xie2014NC,Zhu2016NC,Li2017PRL,Assouar2018NRM,Zhu2021AFM} and shown that lossy acoustic metasurfaces ~\cite{Liu2018PRL,Wang2019PRL,Gerard2020MRS,Song2019PRApplied,Yang2019PRApplied} could serve as a fertile platform for engineering EPs. In particular, it was illustrated that a 2nd order EP derived from a $2 \times 2$ scattering matrix, which portrays a 2-channel metasurface, could give rise to extremely asymmetrical retro-reflection ~\cite{Wang2019PRL}. Extending this concept to higher-order EPs, however, is not trivial, since conventional metasurfaces do not offer precise and simultaneous control over the multiple propagating orders existing in the corresponding higher-order scattering matrices, which is instrumental for achieving higher-order EPs. This hurdle can be attributed to the fact that traditional metasurfaces treat each sub-unit separately without considering their nonlocal interaction, and as a result the diffraction efficiency is fundamentally limited. 
%New routes that overcome this limitation would therefore greatly aid non-Hermitian metasurface-based concepts alongside facilitating the path to higher-order EPs via their elegantly engineered scattering matric

This letter reports on the experimental observation of a higher-order EP in an open acoustic system, and illustrates that the nonlocal metagrating ~\cite{Alu2017PRL,Quan2019PRL,Overvig2020PRL,Torrent2018PRB} is the key enabler for this observation. We start by establishing a mathematical framework to construct an S-matrix that can give rise to arbitrary order EPs. The metagratings that are then proposed comprise lossy and non-lossy sub-units, where the former represents loss in a passive $\mathcal{PT}$ symmetric medium~\cite{Liu2018PRL}. Our nonlocal metagratings can be conveniently tuned to tailor the S-matrix, offering a robust approach for engineering EPs of arbitrary orders. Additionally, the design that is put forward here, harnesses the nonlocal response of the constituent sub-units via evanescent wave fields, thereby offering the ability to efficiently mold the energy flow to multiple channels of wave reflection. This is in stark contrast to the prior designs for nonlocal acoustic wave manipulation ~\cite{Quan2019PRApplied,Quan2019PRL} and higher order EPs ~\cite{Ding2015PRX}, that rely on physical connections between their cavities. Our work illustrates that non-local metagratings could offer greater flexibility for the study of higher-order EPs in open systems while simultaneously enabling unprecedented wavefront engineering capabilities.  

Without loss of generality, a square matrix, A, of arbitrary order, $N$, have its eigenvalues and eigenvectors coalesce simultaneously only when it is a \textit{similarity matrix} of the Jordan-block canonical form, J, which can be expressed as,  
\begin{equation}
{{\text{A}}^{\left( N \right)}} \sim \;{{\text{J}}^{\left( N \right)}}\left( {{E_0}} \right) = \left( {\begin{array}{*{20}{c}}
  {{E_0}}&1&{}&{}&{} \\ 
  {}&{{E_0}}&1&{}&{} \\ 
  {}&{}& \ddots & \ddots &{} \\ 
  {}&{}&{}&{{E_0}}&1 \\ 
  {}&{}&{}&{}&{{E_0}} 
\end{array}} \right).\label{eq:0}
\end{equation}
where $\sim$ is the similarity symbol and $E_0$ is the degenerate eigenvalue. %While several recent studies have explored this coalescence for Hamiltonians with $N$ $\ge$ 2 \cite{Hodaei2017Nature,Ding2015PRX}, EPs derived from higher-order scattering matrices (S-matrices) remain elusive. 
Here, we access the EPs in higher-order S-matrices via a non-local reflective metagrating that offers complete control over the multiple modes of diffraction.

As per the classical diffraction theory, the relationship between the incoming wave and an $n$-th order reflected wave reads ${k_0}\left( {\sin {\theta _r} - \sin {\theta _i}} \right) = nG\label{eq:1} $. Here, $k_0$, is the wavenumber in free-space and $\theta_i$ and $\theta_r$ are the angles of incidence and reflection, respectively. $G = 2\pi/D$ is the reciprocal lattice vector, where $D$ is the grating period that can be tailored to enable different orders of diffraction. Here we assume the operating frequency is 3430 Hz. In the case of the 3rd order system, for example, we require the existence of three propagating channels - $D$ must be $\sqrt{2}$ times the wavelength, $\lambda$, and would result in a $G=(\sqrt{2}k_0)/2$.  The response of the resulting system can be then described as  ${\left\{ {p_r^R,p_r^N,p_r^L} \right\}^{\rm{T}}} = {\rm{S}}{\left\{ {p_i^L,p_i^N,p_i^R} \right\}^{\rm{T}}}$, where the vector on the left denotes the output sound field (reflection) while that on the right denotes the input (incidence). Here, $p$ is the complex pressure amplitude, whose superscript indicates the left (L), normal (N) and right (R) channels of the grating, and the subscript refers to the reflection ($r$) or incidence ($i$). The S-matrix that connects these two vectors can be written as, 
\begin{equation}
{{\rm{S}}} = \left( {\begin{array}{*{20}{c}}
{r_0^L}&{r_{ + 1}^N}&{r_{ + 2}^R}\\
{r_{ - 1}^L}&{r_0^N}&{r_{ + 1}^R}\\
{r_{ - 2}^L}&{r_{ - 1}^N}&{r_0^R}
\end{array}} \right),\label{eq:2}
\end{equation}
where the individual elements denote the reflection coefficient of each mode. The superscripts indicate the directions of incidence ($L$, $N$ and $R$ refer to left  $45^{\circ}$, normal $0^{\circ}$, and right $-45^{\circ}$, respectively) and the subscripts represent the orders of reflection. The non-clinodiagonal terms correspond to the specular ($r_0^L,r_0^R$) and anomalous reflection ($r_{ - 1}^L,r_{ - 1}^N,r_{ + 1}^N,r_{ + 1}^R$) cases, where in both cases the wave gets reflected from a port different from that of the incident one. The clinodiagonal terms ($r_{ - 2}^L,r_0^N,r_{ + 2}^R$) of the matrix denote retro-reflection, where the wave goes back in the same direction from where it is incident. These retroreflection terms are of great interest in this study, since they hold the key to achieving  interesting phenomena at the 3rd order EP that can be experimentally observed.

Here, we focus on an extreme scenario where $r^L_{0}$, $r^N_{0}$, $r^R_{0}$, $r^{N}_{+1}$, $r^{R}_{+1}$, and $r^{R}_{+2}$ all vanish, $r^{L}_{-2}$ is as large as possible, while $r^L_{-1}$ and $r^N_{-1}$  approach zero (they can not be exactly zero). According to linear algebra, the resulting S-matrix is clearly a  similarity transformation of matrix J (with $E_0$ being zero) shown in Eq. \ref{eq:0}, with the change of basis matrix ${\rm{P}} = \left( {\begin{array}{*{20}{c}}
0&0&1\\
0&{r_{ - 1}^L}&0\\
{r_{ - 1}^Lr_{ - 1}^N}&{r_{ - 2}^L}&0
\end{array}} \right)$.  

\begin{figure}
\includegraphics[width=8.6cm]{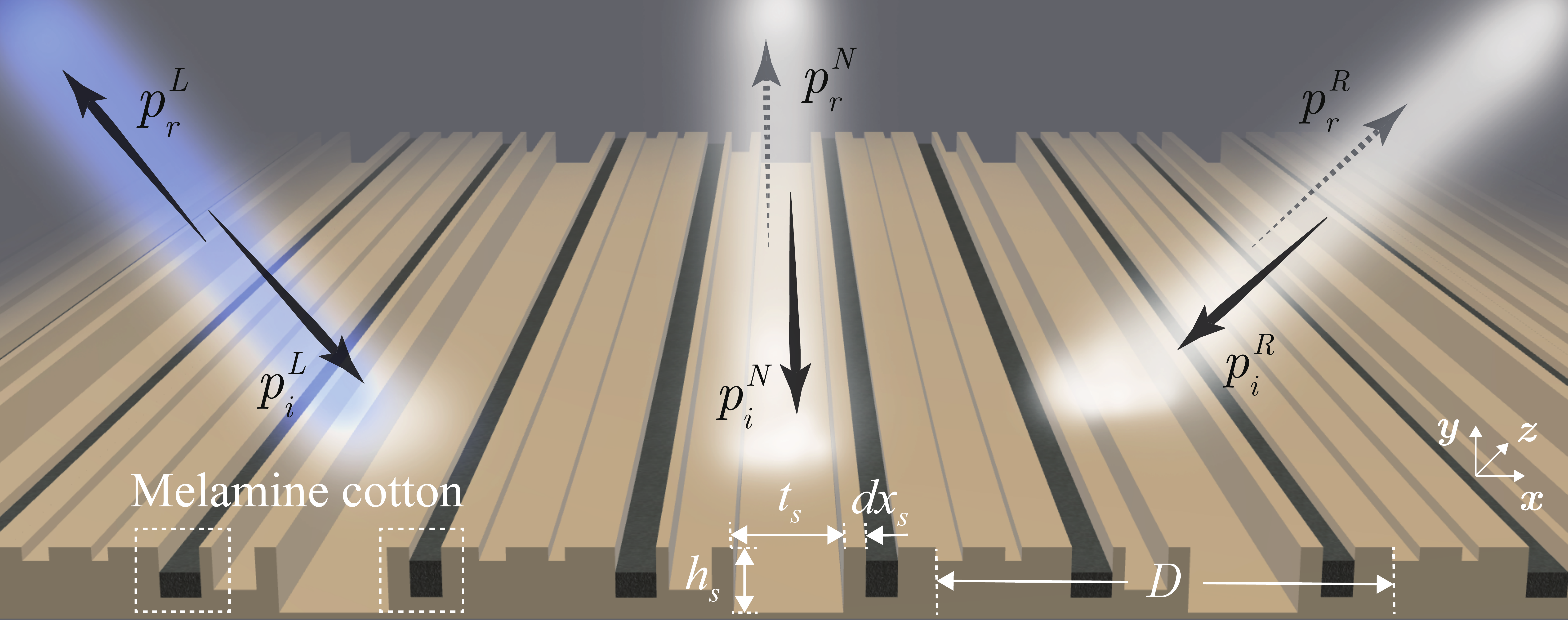}
\caption{\label{fig:1} Schematic diagram of a non-local and non-Hermitian acoustic metagrating at the higher-order exceptional point. This metagrating is composed of six surface-etched grooves in a period ($D$). The global response of the metagrating to the external excitation can be engineered by tuning the
groove depth (${h_s}$), width (${t_s}$), and the spacing (${dx_s}$), where $s$ is the index of groove in a period (${s=1,2,...,6}$). The variation of ${dx_s}$ can be specifically used to tune the non-local interaction between grooves. In each period, three grooves are filled with sound absorbing materials (melamine cotton) to introduce loss.}
\end{figure}

%\rxadd{To make the S-matrix shown in Eq. \ref{eq:2} a similarity transformation of matrix J shown in Eq. \ref{eq:0}, for the sake of producing  a 3rd-order EP, the three elements in the top right corner of Eq. \ref{eq:2} ($r^N_{+1}$,  $r^R_{+1}$ and $r^R_{+2}$) must be simultaneously zero, while the $r^L_{-1}$ and $r^N_{-1}$ terms should be non-zero.}   The other terms in the matrix can be in fact arbitrary, without moving the system away from the 3rd-order degeneracy. Here, we focus on an extreme scenario where $r^L_{0}$, $r^N_{0}$, $r^R_{0}$, $r^{N}_{+1}$, $r^{R}_{+1}$, and $r^{R}_{+2}$ all vanish, $r^{L}_{-2}$ is as large as possible, while $r^L_{-1}$ and $r^N_{-1}$ infinitely approach zero. 
Our design results in an acoustic mirror that strongly retro-reflects the wave that is incident from one direction, but near completely absorbs those incident from the other two directions, as shown in Fig. ~\ref{fig:1}. Note that, we choose to demonstrate the 3rd-order EP with the most asymmetrical wave behavior, while other types of 3rd-order EPs also exist in the three-channel metagrating. See another example of a metagrating at a different 3rd-order EP in Section A of the Supplemental Material \cite{SP}.

To engineer the S-matrix as described above, we design a metagrating that comprises six surface-etched grooves periodically arranged as shown in Fig. \ref{fig:1}. As opposed to conventional metasurfaces that are designed based on a local phase gradient, the metagrating here leverages the interaction between the constituent grooves (sub-units). An analytical model based on the coupled-mode theory is utilized to take this non-local effect into account and to design a highly efficient three-channel metagrating via global optimization (see Section B in Supplemenatary Material \cite{SP}). Additionally, to induce non-Hermicity to the system, the metagrating has three grooves that are lossy in one period. The lossy grooves possess complex effective sound speeds, whose imaginary parts can be conveniently tuned by embedding sound absorbing materials of appropriate thickness within the groove, as shown in Fig. \ref{fig:1}.

\begin{figure*}
\includegraphics[width=16cm]{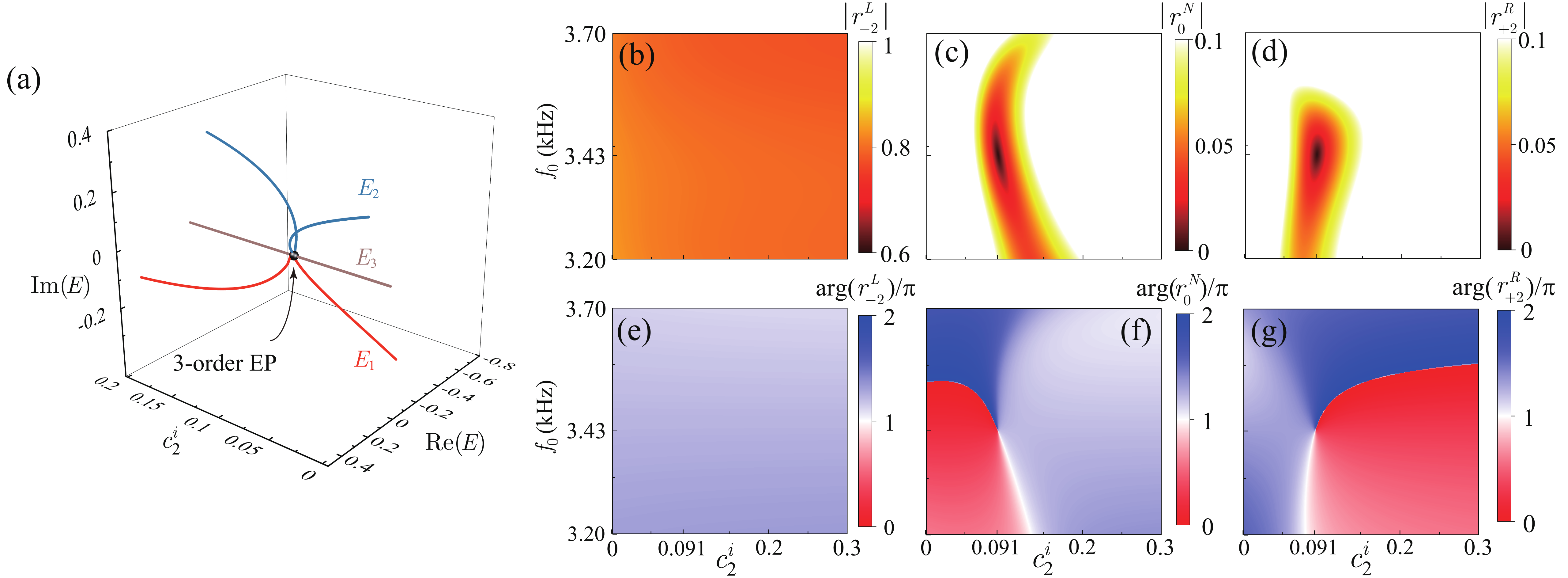}
\caption{\label{fig:2} (a) Variation of trajectories of the eigenvalues with the evolution of ${c_2^i}$. Gray globe in the center of figure represents the $3$-order EP. (b) - (d) Calculated amplitude of retro-reflection coefficients as functions of ${f_0}$ and $c_2^i$ in the left, normal and right channels. $\left| {r_0^N} \right|,\left| {r_{ + 2}^R} \right| \ge 0.1$ inside the white regions in (c) and (d). (e) - (g) Phase of the retro-reflection coefficients around the EP with the variation of ${f_0}$ and $c_2^i$.}
\end{figure*}

As an example, Fig. \ref{fig:2}(a) shows the trajectories of the theoretically calculated eigenvalues as a function of the loss factor of the second groove, $c^i_{2}$ (normalized imaginary part of the sound velocity), while those of the fourth and sixth grooves, $c^i_{4}$ and $c^i_{6}$, are 0.053 and 0.601, respectively. As can be seen here, when $c^i_{2}$ reaches 0.091, the trajectories of the three eigenvalues meet at a crossing point that is a direct manifestation of the 3rd-order EP. In addition, one of the eigenvalues remain stable against the varying loss while the other two vary dramatically in the vicinity of the intersection. This distinctive pattern of the eigenvalues in parameter space is the hallmark of an odd-order EP~\cite{Tang2020Science}.

Figures~\ref{fig:2}(b)-(d) show the amplitudes of the wave that is retro-reflected through the left, normal and right channels, respectively. The three channels respond very differently to the same varying loss. While the left channel retro-reflection remains stable (a high efficiency around 0.81 for the reflection coefficient), the normal and the right channels are highly sensitive to even small variations of the loss factor. As can be seen, the system reaches the 3rd-order EP, when $c^i_{2}$ = 0.091 and $f_0 = 3430 $ Hz, and as a result, the retro-reflections to the right and normal channels vanish, while the one to the left channel still remains strong. Furthermore, the vanishing channels experience a peculiar increase in amplitude beyond the EP, despite the very high loss factor in these regions of parameter space. This peculiar phenomena is known as loss enhanced reflection and is very similar to what was previously observed in a lower-order system \cite{Wang2019PRL}. The phase variation shown in Figs. \ref{fig:2}(e)-(g) also demonstrate the distinct behavior that correspond to the occurrence of the 3rd-order EP: a virtually constant phase shift throughout, for the left channel, but a trip point together with an abrupt phase shift at the EP for the normal and right channels. Furthermore, the amplitude and phase profiles are very much identical to those of the other diagonal coefficients ($r^{L}_{0}$,$r^{R}_{0}$), as well as $r^{R}_{+1}$ and $r^{N}_{+1}$ (Section C of the Supplemental Material \cite{SP}).  

\begin{figure}
\includegraphics[width=7.6cm]{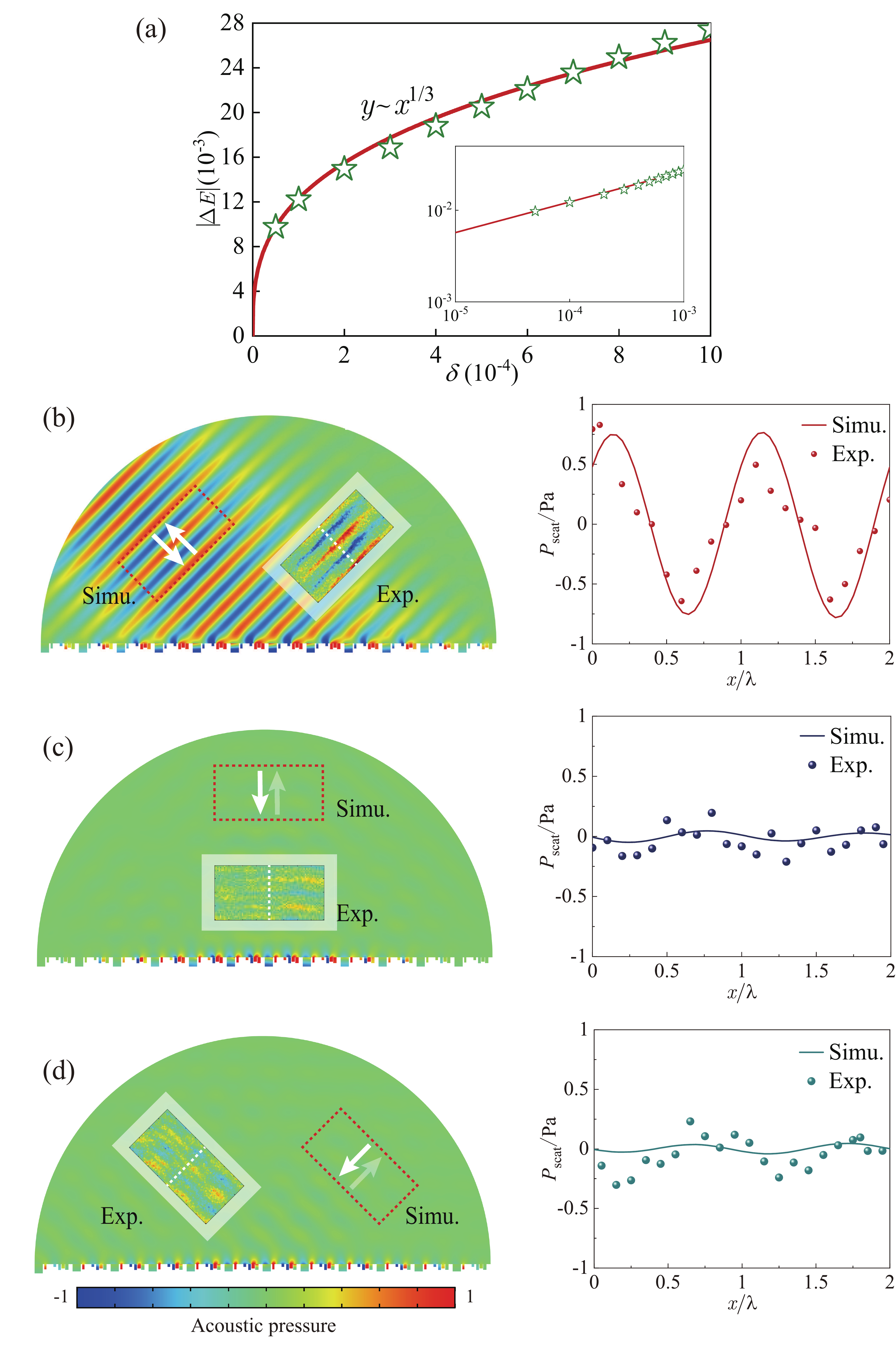}
\caption{\label{fig:3} (a) Sensitivity of the designed metagrating at the 3rd-order EP. $ {\Delta E} $ on $y$-axis represents the difference between real parts of the first and second eigenvalues varying with the external perturbation $\delta $, i.e. disturbances in $c_2^i$. The green stars represent the difference calculated by the coupled mode theory.  Red solid line shows corresponding fitting curves. The inset confirms the variation trajectory in the logarithmic scale. Simulated and experimental results (reflected field) of the multi-channel metagrating at the 3rd-order EP under (b) left, (c) normal, and (d) right incidences. In each case, the scanning region is marked by a dotted-line box in the simulated wave field, and, for comparison, the measured reflected fields are shown with white borders on the side. The right panel shows the simulated (lines) and measured (dots) pressure distributions along the centre line of the scan area, as presented with white dotted lines in the left panel.}
\end{figure}

The strong parameter dependence that is seen here is rather intriguing and can be of great importance for detectors that require high sensitivity. Prior studies ~\cite{Hodaei2017Nature} have shown that the sensitivity, $\Delta$, of a non-Hermitian system would increase with the order of the EP, as $\Delta = \delta^{1/N}$, where $\delta$ represents the external perturbation. To corroborate this trend in the case of nonlocal metagratings, a  2-channel metagrating that features a 2nd-order EP is designed as shown in Section D of the Supplemental Material~\cite{SP}, and its sensitivity is compared with that of the 3rd-order metagrating discussed here. The external perturbation is induced in the lossy sub-unit as a disturbance in $c^{i}_{2}$. The green stars in Fig. \ref{fig:3}(a) show results from the analytical model and the curve-fitted solid red line illustrates that the sensitivity of the designed 3-channel metagrating follows a cube-root dependence on the induced perturbation ($\Delta \propto \delta^{1/3}$). The 2nd order grating on the other hand has $\Delta = \delta^{1/2}$, as shown in Fig. S4 of the Supplemental Marerial \cite{SP}). 
 
\begin{figure*}
\includegraphics[width=16cm]{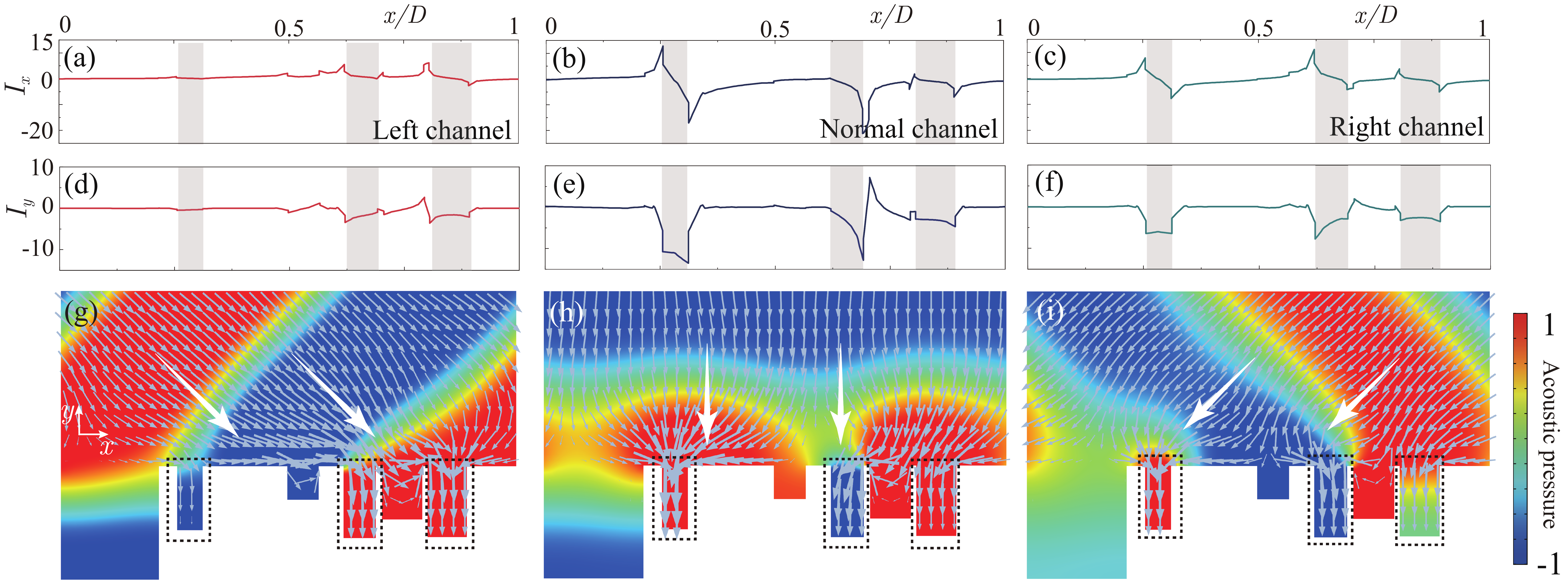}
\caption{\label{fig:4} Non-local intensity flow distribution. (a) - (c) The $x$- and  (d) - (e) $y$-components of the sound intensity (normalized by the incident field) along the metagrating surface under left, normal and right incidences. The regions in grey identify the locations of lossy grooves. (g) - (i) 
The sound field distributions near the surface of the metagrating under left, normal and right incidences, where the color maps illustrate the acoustic pressure distributions and the arrows represent the energy flow. Areas bounded by dotted lines represent the lossy grooves.}
\end{figure*}
 
  The proposed metagrating is then experimentally verified via a fabricated sample of length 1.42 m (10 periods) that was placed in a two-dimensional waveguide of height 4 cm, where a 0.62 m line array with 17 speakers was used as the source. To scan the field, a microphone moved with a step size of 0.01 m over the regions (0.4 $ \times$  0.2 $m^{2}$) that correspond to the three channels, marked by the dotted lines in Fig. \ref{fig:3}. The scattered field was obtained by subtracting the incident field (scanned without the metagrating) from the total field (scanned with the metagrating). Figs. \ref{fig:3}(b)-(d) show the results from numerical simulations carried out by COMSOL Multiphysics, in comparison to the measured results (normalized by the incident pressure) and indicates good agreement. Importantly, the strongly asymmetrical wave behavior is observable in experiments. When the left channel is excited, retro-reflection is clearly observable, and the estimated reflection coefficient reaches around 0.75 (from the simulation and measured data in the right inset of Fig. \ref{fig:3}(b)). When sound is incident on the normal and right channels, however, the retro-reflection is strongly suppressed, as seen in Figs. \ref{fig:3}(c) and (d). Additional details about the metagrating prototype and the experimental platform can be found in Section E of the Supplemental Material \cite{SP}. 
 
 To shed light on the mechanism that gives rise to the asymmetric behavior in this nonlocal non-Hermitian metagrating, the local intensity distribution of the sound field is numerically calculated by COMSOL. Figures~\ref{fig:4}(a)-(c) show the local acoustic intensity in the $x$-direction (${I_x} = \frac{1}{2}{\mathop{\rm Re}\nolimits} \left[ {p\times{{\left( {{v_x}} \right)}^ * }} \right]$, i.e., the product of pressure and complex conjugate of the local velocity in the $x$-direction) at the surface of metagrating for the cases of left, normal, and right incidence cases, respectively. Here, the ``lossy'' sites (three grooves with absorbing layers) are denoted by the regions in grey. It can be observed that $I_x$ changes abruptly at the boundaries of these ``lossy'' regions, manifesting a strong nonlocality-induced lateral energy exchange with the neighboring grooves. Figures~\ref{fig:4}(d)-(f) show the intensity in the $y$-direction (${I_y} = \frac{1}{2}{\mathop{\rm Re}\nolimits} \left[ {p\times{{\left( {{v_y}} \right)}^ * }} \right] $), and illustrate that $I_y$ is negative in the lossy regions due to absorption. It can be noted that in contrast to the other two cases, the intensity fluctuates only slightly for the left incidence case (Figs.~\ref{fig:4}(a) and (d)) and suggests that the lossy grooves respond weakly and therefore renders a stable, highly efficient retro-reflection. This is in accordance with what is observed in Fig.~\ref{fig:2}(b). Figures~\ref{fig:4}(g)-(i) show the energy flow fields, where green arrows in the figures indicate the direction of the local intensity, and the length of arrows represent the intensity amplitude. It can be found that the directions of the arrows are strongly distorted in the area very close to the surface, in sharp contrast to those in the far field. This unveils the fact that empowered by nonlocality, the designed metagrating utilizes the evanescent fields as a mechanism for channeling the energy along the surface of its sub-units.

To conclude, we have theoretically and experimentally investigated a nonlocal acoustic metagrating to illustrate the asymmetrical wave behavior that exists at the higher-order EP of its S-matrix. %In comparison to prior designs for higher-order EPs, nonlocal metagratings are relatively simple since they offer coupling between their constituent units without the need for physical connections among them. Additionally, the study here shows that undertaking such an approach would allow us to marry the rich higher-order EP physics with unprecedented wavefront shaping functionalities. 
Since the general condition for engineering S-matrices with EPs of arbitrary orders is now established, more complicated wave behavior can be envisioned (see the 4th-order EP shown in section F of the Supplemental Material \cite{SP}). Our design suggests an efficient approach towards extremely asymmetric multi-channel wave manipulation with a high and controllable sensitivity, and more essentially, paves  the way for acoustic impedance engineering via multi-functional anisotropic acoustic devices.

\begin{acknowledgements}
	This work was supported by the National Natural Science Foundation of China (Grant Nos. 12074286, 11774265 and 12074288), and the Shanghai Science and Technology Committee (Grant Nos. 20ZR1460900 and 20ZR1461700). N.J.G, Y.D., and Y.J. would like to thank NSF for the support through CMMI 1951221.
\end{acknowledgements}

%\bibliography{bibtext}
%merlin.mbs apsrev4-1.bst 2010-07-25 4.21a (PWD, AO, DPC) hacked
%Control: key (0)
%Control: author (8) initials jnrlst
%Control: editor formatted (1) identically to author
%Control: production of article title (-1) disabled
%Control: page (0) single
%Control: year (1) truncated
%Control: production of eprint (0) enabled
%

\end{document}